\documentstyle[aps,prl,graphicx,amsbsy]{revtex}
\begin{document}
\wideabs{

\title{First order phase transition from the vortex liquid to an amorphous
solid}
\author{M. Menghini$^1$, Yanina Fasano$^1$, F. de la Cruz$^1$, S.S. Banerjee$^2$, Y. Myasoedov$^2$,
E. Zeldov$^2$, C. J. van der Beek$^3$, M. Konczykowski$^3$ and T.
Tamegai$^4$}

\address{$^1$Instituto Balseiro and Centro At\'{o}mico Bariloche, CNEA, Bariloche, 8400, Argentina}
\address{$^2$Department of Condensed Matter Physics,
Weizmann Institute of Science, Rehovot 76100, Israel}
\address{$^3$Laboratoire des Solides Irradi\'{e}s, CNRS UMR 7642,
Ecole Polytechnique, 91128 Palaiseau, France}
\address{$^4$Department of Applied Physics, The University of Tokyo, Hongo, Bunkyo-ku,
Tokyo 113-8656, and CREST Japan Science and Technology Corporation
(JST), Japan}

\date{\today}
\maketitle

\begin{abstract}
We present a systematic study of the topology of the vortex solid
phase in superconducting Bi$_{2}$Sr$_{2}$CaCu$_{2}$O$_{8}$ samples
with  low doses of columnar defects. A new state of vortex matter
imposed by the presence of  geometrical contours associated with
the random distribution of columns is found. The results show that
the first order liquid-solid  transition in this vortex matter
does not require a structural symmetry change.

\end{abstract}

\pacs{PACS 74.60.Ge; 74.60.Dh; 74.60.Jg; 74.25.Bt; 74.25.Ha;
74.72.Ny}

}

As a consequence of the interaction between vortices and quenched
disorder in superconducting samples, the vortex liquid solidifies
through a first order phase transition (FOT)
\cite{safar,hernan,zeldov} into a topologically ordered solid with
elastic deformations, the Bragg glass
phase\cite{giamarchi,neutrones}. Artificially introduced
correlated defects, as those produced by heavy ion irradiation,
transform the first order liquid-solid transition into a second
order one in which the solid vortex phase, the Bose
Glass\cite{nelson}, has no topological quasi long range order.
However, studies \cite{kaykovich1} of the phase diagram of
Bi$_{2}$Sr$_{2}$CaCu$_{2}$O$_{8}$ (BSCCO-2212) have shown that the
FOT is robust in the presence of low density of columnar defects.
In this case the FOT is preserved even for vortex densities
comparable to those of defects.

High dose of  columnar defects has two main effects on the vortex
system. On one hand, its random distribution suppresses the FOT;
on the other hand, the correlated potential hardens the vortex
lines inducing a divergent vortex tilting elastic constant,
$C_{44}$, at the Bose Glass transition. The results in Ref.
\onlinecite{kaykovich1}  indicate that the disorder induced by a
low density of columnar defects is too weak to change the order of
the liquid-solid transition.  This could imply that the quasi long
range order of the Bragg Glass is preserved despite the presence
of correlated defects.

The aim of this work is to establish the correlation between the
space configuration of columnar defects and the vortex structure
created in the presence of those defects. This goal is achieved
visualizing the vortex structure  by means of the magnetic
decoration technique in samples where the liquid-solid phase
transition has been characterized by differential magneto optics
(DMO) \cite{soibel}. The observed vortex topology is discussed in
terms of the nature of  the liquid-solid transition.

The BSCCO-2212 samples used in this work were irradiated at GANIL
by 1 GeV Pb ions parallel to the $c$ axis of the crystal. In order
to compare the vortex structure in the presence of  columnar
defects with that of  the pristine sample, the irradiation was
made through an ad-hoc stainless steel mask that allows to have
irradiated and non irradiated regions in the same sample (see
inset of Fig\ \ref{fig:1}(a)). We have investigated four samples
irradiated with doses corresponding to $B_{\Phi} = 5, 10, 50$ and
$100 \,$G, where $B_{\Phi}$, the matching field, is usually
defined as $n_{col}\Phi_{0}$ with $n_{col}$  the density of
columns and $\Phi_{0}$  the flux quantum. The $H-T$ phase diagram
obtained by DMO imaging \cite{porous} for these samples is shown
in Fig.\ \ref{fig:1}(a). It is important to remark that this high
sensitivity technique shows that the FOT is not only preserved in
the presence of  low density of columnar defects but the
transition temperature is raised when compared to that of pristine
samples. For samples with $B_\Phi =5$ and $10\,$G the melting line
is first order and for the $B_\Phi =100\,$G sample is continuous
in the range of fields investigated by magnetic decoration. For
the $B_\Phi =50\,$G sample  a critical point at $70\,$G separating
the FOT at lower fields  from a continuous one at higher fields is
found. The  magnetic decoration was performed after
cooling the sample in the presence of a field, FC experiment,
down to $4.1\,$K. The results obtained in the non irradiated
regions confirm the high quality of the crystals.

Figure\ \ref{fig:1}(b) shows the vortex structure in the non
irradiated and irradiated regions obtained by magnetic decoration
after FC the  $B_{\Phi}=10\,$G sample in a field of  $40\,$G. A
sharp boundary separates the Bragg Glass  in the non irradiated
region from a structure with no long range crystalline order in
the irradiated one.

The experimental results show that the vortex structure is polycrystalline for
$B>B_{\Phi}$ and amorphous in the range $B<B_{\Phi}$. Examples of Delaunay
triangulations of polycrystalline vortex structures observed by magnetic
decoration after FC the $B_{\Phi}=10\,$G sample in fields of  $80\,$G and
$40\,$G are shown in Fig.\ \ref{fig:2}(a) and (b), respectively. Fig.\
\ref{fig:2}(c) shows the vortex structure for $B=B_{\Phi}$. The black dots
indicate topological defects (non-sixfold coordinated vortices). Figure\
\ref{fig:3} depicts decoration images of amorphous structures obtained in the
regime $B<B_{\Phi}$; (a) after FC the $B_{\Phi}=50\,$G sample in a field of
$30\,$G and (b) after FC the $B_{\Phi}=100\,$G sample in a field of $80\,$G.

Contrary to what is often observed in homogeneous materials, we argue that the
nature of the  polycrystalline structure shown in Fig.\ \ref{fig:2} is an
intrinsic property of the vortex matter in the presence of randomly distributed
correlated defects. To demonstrate this we have used the dynamic annealing
method \cite{prb62,peakII} in which the magnetic field is tilted back and forth
before the decoration is made. This procedure is known to remove the grain
boundaries in NbSe$_{2}$ and to reorder the crystalline vortex lattice
\cite{prb62} in non irradiated BSCCO-2212 samples. In contrast, we find that
the polycrystalline structure of the irradiated region  is not altered after
applying this annealing procedure. This supports that the observed structure is
the equilibrium one rather than a metastable configuration resulting from the
typical nucleation and growth processes.

The above discussion suggests that the grain boundaries of the
polycrystalline structure are fixed in space  by the random
distribution of columnar defects. If this were the case the area
of the grains for a given $B_{\Phi}$ would be field independent
contrary to what is observed in homogeneous low $T_c$
superconductors. To verify this, we studied the statistical
distribution of the grain areas as a function of the applied field
for $B_{\Phi}=5$ and $10\,$G samples.  The grain boundaries are
defined as the interface between crystallites which orientations
differ by 10 degrees or more. The areas of the different grains
are measured counting the number of vortices within them. The
analysis of many pictures shows that the histogram of the grain
sizes is broad with upper and lower limiting values for the grain
areas for each applied field. The number of vortices, $N_{vg}$,
within the largest and smallest grains (upper and lower limits of
the histogram) are proportional to $B$  as shown in Fig.\
\ref{fig:4} for $B_{\Phi}=10\,$G sample. Identical behavior is
found in the $B_{\Phi}=5\,$G sample with the shift of the
distribution of grain areas corresponding to the lower irradiation
dose. Furthermore, the average number of vortices per grain
(defined as the ratio between the total number of vortices within
grains and the number of grains) is also found to be proportional
to $B$, see Fig.\ \ref{fig:4} (solid squares). These results  make
evident that the size of the different grains for a given
$B_{\Phi}$ is field independent contrary to what is found in low
$T_c$ superconductors, where the average grain size grows with the
magnetic field \cite{marchevsky}. The invariance of the grain size
with $B$ in the irradiated regions strongly supports that the
grain size as well as the space distribution of the grains are
determined by the landscape of the columnar defects.

In order to understand the relation between grain boundaries and the
distribution of correlated defects we have generated a set of random points
simulating the positions of columnar defects, see Fig.\ \ref{fig:2}(d) (white
dots). This configuration is found to be equivalent to that of columnar defects
detected after etching mica irradiated by heavy ions with comparable doses, see
inset in Fig.\ \ref{fig:4}. It is interesting to remark that the distribution
of columnar defects is not homogeneous on the scale of the lattice parameter in
the range of fields investigated. The in-plane inhomogeneous distribution and
the correlated character of the columnar defects suggest the presence of a
network of contours extending throughout the thickness of the sample associated
with the localization of vortices on columns. In Fig.\ \ref{fig:2}(d) the
contours are represented by the gray connected areas indicating the range of
interaction ( penetration depth) of vortices localized on defects. These
vortices create energy barriers that inhibit the propagation of the crystalline
vortex lattice, giving rise to the polycrystalline structure observed in the
regime $B>B_{\Phi}$. Thus, the observed grain boundaries in samples with low
irradiation dose have a different origin than those associated with metaestable
states in homogeneous systems \cite{peakII}. The apparent grain boundaries in
the irradiated samples are field independent vortex contours fixed in space by
the random distribution of columns. This prompts us to propose a new state of
vortex matter where the total number of vortices is divided in two species, one
corresponds to the fraction of vortices localized on contours, $\rho_{cont}$,
and the other forming the crystallites surrounded by the contours,
$\rho_{crys}$.

Based on the above discussion we present a simplified picture that allows to
predict the dependence of the number of topological defects (non sixfold
coordinated vortices) of the polycrystalline structure, $N_{def}$, on $B$ and
$B_{\Phi}$. The total number of defects is counted in an arbitrary large area
that contains many grains. The reported data was obtained from an inspected
area of the order of $10^4\,\mu\,m^2$, much larger than the maximum grain size
for the different $B_{\Phi}$ investigated. From the inspection of many images
of the vortex structure in the irradiated regions we find that most of the
topological defects are located along the contours separating the crystallites.
Thus, as a first order approximation we assume that $N_{def}$ is proportional
to the number of vortices localized on the contours, $N_{cont}$. The total
length of the contours within the inspected area can be written as
$L=N_{cont}\,a_0$ where $a_0$ is the average vortex distance. The experimental
results showing the field independence of the areas of the crystallites (see
Fig. \ \ref{fig:4}) imply that $L$ is only function of $B_{\Phi}$. Therefore
$N_{def}\propto\,N_{cont}\propto\,L(B_{\Phi})/a_{0}\propto\,L(B_{\Phi})B^{1/2}$.
On the other hand, the self similarity of the space distribution of random
columnar defects imposes that the areas enclosed by the contours scale with
$B_{\Phi}$ and consequently $L(B_{\Phi})\propto B_{\Phi}^{1/2}$. Therefore, the
fraction of vortices involved in topological defects, $\rho_{def}$, is

\begin{equation}
\rho_{def}=N_{def}/N_v\propto(B_{\Phi}/B)^{1/2} \label{eq1}
\end{equation}

In Fig.\ \ref{fig:5} we plot $\rho_{def}$ obtained from Delaunay triangulations
of vortex structures for all $B$ and $B_{\Phi}$ investigated. The black line
corresponds to the square root dependence predicted by Eq. \ref{eq1}. Despite
the crude approximations, $N_{def}\propto N_{cont}$ and $L(B_{\Phi})\propto
B_{\Phi}^{1/2}$, we see that for $B_{\Phi}<B$ the experimental results are well
described by this simple model. On the other hand, the deviation of the
experimental data from the dependence predicted for $\rho_{def}$ in the range
$B_{\Phi}/B>1$ is evident.

The failure of the model to describe the amorphous state is
expected, since for fields $B \leq B_{\Phi}$  a significative
fraction  of the areas enclosed by the contours fixed in space are
not large enough to allocate vortices with crystalline symmetry,
$\rho_{crys}\approx0$. In this limit, despite most of the vortices
remain pinned on columns, $\rho_{cont}\approx1$, the number of
topological defects  is not proportional to $L(B_{\Phi})/a_0$;
i.e. $a_0$ is not an appropriate length scale to apply the scaling
argument.

The $B_{\Phi}/B$ independence of the fraction of defects
 in the amorphous state ($\rho_{def}\approx 0.5$) is
remarkable, see Fig.\ \ref{fig:5}. To investigate the origin of this result we
have generated a random distribution of points with the constrain that the
distance between near neighbors is not less than $50\%$ of the lattice
parameter of a perfect lattice with the same density. The constrain is chosen
to allow the fluctuations of the average distance between points to be that
observed in the amorphous vortex structure. It is found that the fraction of
defects obtained from Delaunay triangulation of this simulated distribution of
points is the same as that for the vortex structure for $B<B_{\Phi}$,
indicating that in this regime most of the vortices are pinned on a random
distribution of columnar defects.

The analysis of the vortex structure for $B>B_{\Phi}$ as well as
for $B<B_{\Phi}$, Fig.\ \ref{fig:5}, strongly supports that in
both limits the topological defects are associated with vortices
localized  on columns. For $B>B_{\Phi}$ the topological defects of
the vortex structure are mainly due to plastic distortions induced
by the presence of crystallites. In the other limit, $B<B_{\Phi}$,
the space within the contours does not allow to form crystallites,
$\rho_{crys}\approx 0$, and the defects are associated with
vortices localized on a fraction of columns compatible with the
constrain mentioned above. It is clear that $B/B_{\Phi}\approx1$
marks the transition from a topological state described by a
collection of crystallites to a state characterized by an
amorphous structure, both determined by the random distribution of
columns.

As a consequence of the results discussed above we see that in the
limit  $B<B_{\Phi}$  no crystalline structure can be accommodated
within the small contours determined by the columnar defects,
$\rho_{crys}\approx 0$. Thus, an amorphous vortex structure
characterizes the solid state. This is particularly relevant when
analyzing the interrelation between the solid structure and the
nature of the vortex liquid-solid transition. The critical point
at $70\,$G  in the phase diagram of the $B_{\Phi}=50\,$G sample
indicates that  most of the field range where the first order
melting takes place corresponds to the regime $B<B_{\Phi}$. Thus,
the observed amorphous vortex structure, see Fig.\ \ref{fig:3}(a),
rules out the widely accepted correlation between a  first order
liquid-solid transition and a structural symmetry transformation.
Moreover, we observe no difference between the topological order
in Figs.\ \ref{fig:3} (a) and (b) despite the fact that the former
structure has solidified through a FOT whereas the latter through
a continuous phase transition.

In summary, a new vortex matter with two types of vortices has been discovered,
giving support to the recently suggested vortex porous structure \cite{porous}
and the interstitial liquid\cite{radzihovsky}. One type is associated with
vortices localized upon the topological contours formed by the random
distribution of columnar defects and the other with vortices forming the
crystallites. The relative distribution of vortices in these two classes
explains the topological transformation of the solid from a Bragg Glass
($B_{\Phi}=0$) into a polycrystal ($B_{\Phi}<B$) and then into an amorphous
structure ($B_{\Phi}>B$) as the density of columnar defects is increased. The
lack of correlation between the structure of the vortex solid and the order of
the melting transition opens an important question on the microscopic mechanism
that triggers melting in vortex solids. The detection of a FOT between an
amorphous solid and a liquid makes evident that the first order melting
transition in vortex matter does not require the widely accepted long range
structural order of the solid state.

MM, YF and FC acknowledge E. Jagla, P. Cornaglia and F. Laguna for
useful discussions and suggestions. FC and EZ acknowledge the
support by the Fundaci\'on Antorchas-WIS collaboration program.
Work partially supported by ANPCYT PICT99-5117.  M. M. and Y. F.
hold a fellowship of CONICET.

\begin{figure}[hhh]
%\vspace{-5mm}
%\hspace{-5mm}
%\includegraphics[]{fig1.ps}
\caption[]{(a) Phase diagram obtained by DMO for the pristine
BSCCO-2212 sample and for samples with columnar defect  densities
corresponding to $ B_{\Phi}= 5,10,50$ and $100\,$G.  Solid lines
indicate first order liquid-solid  transition while dotted lines
correspond to continuous ones. Inset: Schematics of the  sample
with the mask used for irradiation. (b) Magnetic decoration at
$40\,$G  in the sample with $B_{\Phi}= 10\,$G. The dashed line
shows the boundary between the irradiated and the non-irradiated
regions.} \label{fig:1}
\end{figure}

\begin{figure}[hhh]
%\vspace{-5mm}
%\hspace{-5mm}
%\includegraphics[]{fig2.ps}
\caption[]{Delaunay triangulations of the vortex structure in a
sample with $B_{\Phi}= 10\,G$ at fields of: (a) 80 G, (b) 40G and
(c) 10 G. The black dots indicate the non sixfold coordinated
vortices. The gray lines in (a) and (b) depict the grain
boundaries. (d) Set of random points (white dots) generated
numerically simulating a distribution of columnar defects  with
$B_{\Phi}=10\,$G. The gray areas indicate the range of interaction
of vortices localized on defects, see text.} \label{fig:2}
\end{figure}

\begin{figure}[hhh]
%\vspace{-5mm}
%\hspace{-5mm}
%\includegraphics[]{fig3.ps}
\caption[]{(a) Structure of the vortex solid after a FOT  at
$B=30\,$G   in the $B_{\Phi}=50\,$G sample. (b) Structure of the
vortex solid after a continuous liquid-solid transition at 80 G in
the $B_{\Phi}=100\,$G sample.} \label{fig:3}
\end{figure}

\begin{figure}[hhh]
%\vspace{-5mm}
%\hspace{-5mm}
%\includegraphics[]{fig4.ps}
\caption[]{Number of vortices within the largest (triangles), the
average (squares) and the smallest (circles) grains as a function
of  $B$ for $B_{\Phi}=10\,$G. Inset: Columnar defect distribution
in mica irradiated with heavy ions.} \label{fig:4}
\end{figure}

\begin{figure}[hhh]
%\vspace{-5mm}
%\hspace{-5mm}
%\includegraphics[]{fig5.ps}
\caption[]{Fraction of defects, $\rho_{def}$ as a function of
$B_{\Phi}/B$ for all $B_{\Phi}$ investigated. The black line
corresponds to $\rho_{def}\propto(B_{\Phi}/B)^{1/2}$ and the gray
dotted line to $ \rho_{def}=0.5$.} \label{fig:5}
\end{figure}

\end{document}